\documentclass[aps,prl,twocolumn,10pt,amsmath,amssymb,bibnotes,superscriptaddress,longbibliography]{revtex4-1}

\usepackage{graphicx}
\usepackage{siunitx}
\usepackage{braket}
\usepackage[colorlinks,urlcolor=blue,citecolor=blue,linkcolor=blue]{hyperref}
\usepackage[normalem]{ulem}

\newcommand{\ics}{\ensuremath{I_{CS}}}
\newcommand{\kb}{\ensuremath{k_\text{B}}}

\newcommand{\vptwo}{\ensuremath{V_{P2}}}
\newcommand{\vpone}{\ensuremath{V_{P1}}}
\newcommand{\n}{\ensuremath{N_{1}}}
\newcommand{\vd}{\ensuremath{\tilde{V}_{D}}}
\newcommand{\eps}{\ensuremath{\epsilon_{1}}}

\begin{document}

\title{Remote entropy measurement in coupled quantum dots}  \author{Owen Sheekey}
	\affiliation{Stewart Blusson Quantum Matter Institute, University of British Columbia, Vancouver, British Columbia, V6T1Z4, Canada}
	\affiliation{Department of Physics and Astronomy, University of British Columbia, Vancouver, British Columbia, V6T1Z1, Canada}
\author{Tim Child}
	\affiliation{Stewart Blusson Quantum Matter Institute, University of British Columbia, Vancouver, British Columbia, V6T1Z4, Canada}
	\affiliation{Department of Physics and Astronomy, University of British Columbia, Vancouver, British Columbia, V6T1Z1, Canada}
\author{Elena Cornick}
	\affiliation{Stewart Blusson Quantum Matter Institute, University of British Columbia, Vancouver, British Columbia, V6T1Z4, Canada}
	\affiliation{Department of Physics and Astronomy, University of British Columbia, Vancouver, British Columbia, V6T1Z1, Canada}
\author{Saeed Fallahi}
	\affiliation{Department of Physics and Astronomy, Purdue University, West Lafayette, Indiana, USA}

\author{Geoffrey C. Gardner}
	\affiliation{Microsoft Quantum, West Lafayette, Indiana, USA}

\author{Michael J. Manfra}
	\affiliation{Department of Physics and Astronomy, Purdue University, West Lafayette, Indiana, USA}
     \affiliation{Microsoft Quantum, West Lafayette, Indiana, USA}
    \affiliation{Elmore Family School of Electrical and Computer Engineering, Purdue University, West Lafayette, Indiana, USA}
    \affiliation{School of Materials Engineering, Purdue University, West Lafayette, Indiana, USA}
    \affiliation{Purdue Quantum Science and Engineering Institute, Purdue University, West Lafayette, Indiana, USA}

\author{Eran Sela}
	\affiliation{School of Physics and Astronomy, Tel Aviv University, Tel Aviv 6997801, Israel}
\author{Yaakov Kleeorin}
	\affiliation{Center  for  the  Physics  of  Evolving  Systems,  Department of Biochemistry and Molecular Biology, University  of  Chicago,  Chicago,  IL,  60637,  USA}
\author{Yigal Meir}
	\affiliation{Department of Physics, Ben-Gurion University of the Negev, Beer Sheva 84105, Israel}
\author{Silvia L\"{u}scher}
\email{luescher@physics.ubc.ca}
	\affiliation{Stewart Blusson Quantum Matter Institute, University of British Columbia, Vancouver, British Columbia, V6T1Z4, Canada}
	\affiliation{Department of Physics and Astronomy, University of British Columbia, Vancouver, British Columbia, V6T1Z1, Canada}
\author{Joshua Folk}
\email{jfolk@physics.ubc.ca}
	\affiliation{Stewart Blusson Quantum Matter Institute, University of British Columbia, Vancouver, British Columbia, V6T1Z4, Canada}
	\affiliation{Department of Physics and Astronomy, University of British Columbia, Vancouver, British Columbia, V6T1Z1, Canada}
\date{\today}

\begin{abstract}
Recent experiments have demonstrated that measurements of the entropy change associated with the addition of electrons to semiconductor- and graphene-based quantum dots accurately quantify the spin and orbital degeneracy of the states into which they are added.
However, measuring more exotic entropies requires probing the entropy change of an entire system in response to an added particle. Here, we demonstrate that Maxwell relation-based measurements probe not only the entropy change associated with the added electron but also that of the surrounding system as it responds to that electron.
Using a pair of capacitively coupled GaAs quantum dots, we show that charge measurements on one dot reveal entropy changes associated with the entire two-dot system, both at weak dot--reservoir coupling where microstate counting applies and at stronger coupling where numerical renormalization group calculations are required.
\end{abstract}

\maketitle

 Entropy has been proposed as a powerful tool to identify non-trivial electronic states in mesoscopic systems, from Majorana zero modes \cite{Sela.2019} or non-abelian quasiparticles in the fractional quantum Hall regime \cite{BenShach.2013, Cooper.2008}, to topological entanglement entropy of fractional quantum Hall edges~\cite{sankar2023measuring}. Within the last few years, experiments have demonstrated that thermodynamic Maxwell relations can indeed be employed to measure the entropy of quasiparticles in a variety of low-dimensional systems, including twisted bilayer graphene \cite{Rozen.2021, saito2021isospin}, single and double quantum dots (QDs)~\cite{Hartman.2018, child2022robust, child2022entropy, Adam2025Entropy,Kealhofer2025Entropy}, and even single molecules~\cite{Pascal.2021,pyurbeeva2021controlling}.  In each of these experiments, gate voltages were used to add electrons to the system in question, then the entropy change resulting from electron addition was extracted via a Maxwell relation that connects entropy with particle number, chemical potential, and temperature.

The entropy change reflected in the Maxwell relation is that of the entire thermodynamic system in equilibrium with the reservoir, which may include elements beyond the subsystem to which the electron is added.  In previous experiments, however, the rest of the system was unaffected by the gate voltage, so the measured entropy change could be attributed directly to the local degrees of freedom of the added electron: in Ref.~\cite{Hartman.2018}, for example, sweeping a gate voltage across the $0\to 1$ transition yielded simply the spin entropy of the added electron.

Measuring more exotic entropies---for example, the $k_B\ln\sqrt{2}$ entropy of a Majorana zero mode~\cite{Sela.2019}, or the topological entanglement entropy of a quantum Hall edge~\cite{sankar2023measuring}---requires a different approach, in which the dot receiving the electron serves as an auxiliary control element whose occupation changes the state of a separate part of the circuit.  The entropy measured via $dN/dT$ on the auxiliary dot then reflects not only the particle's own degrees of freedom, but the response of the broader system to its addition.

Considering a specific example, Ref.~\cite{Sela.2019} proposes to measure the entropy of a Majorana zero mode in a proximitized nanowire by the simple addition of a spinless electron ($S$=0) to a nearby QD. If the QD is electrostatically coupled to the barrier at the end of the wire, an extra electron in the dot can quench the hybridization of the Majorana zero mode with a normal metal lead, thereby making the entropy of the Majorana zero mode visible in a low temperature measurement.  The Maxwell relation then implies that the temperature dependence of the QD occupation would reflect the entropy of the Majorana zero mode: the QD is effectively a `remote' entropy sensor for the Majorana zero mode in the nanowire.

Here we test this idea experimentally in a pair of capacitively coupled GaAs quantum dots (QD1 and QD2) \cite{chan2002strongly, hubel2007two, PhysRevLett.101.186804}, a system known to host a variety of strongly correlated ground states under appropriate tuning \cite{borda20034, galpin2005quantum, mitchell2006gate, PhysRevB.84.161305, ferreira2011capacitively, PhysRevLett.110.046604, keller2014emergent, krychowski2016spin, hou2020many, lombardo2020kondo}. Charge measurements on QD1 resolve entropy changes of the full two-dot system, first in the weakly-coupled regime where the microstate count is unambiguous, and then as the dot-reservoir coupling becomes strong.

Our protocol leverages a slightly different version of the more common Maxwell relation\cite{child2022entropy} 
\begin{equation}\label{eq:eq1}
    \left(\partial S_{sys}/ \partial \epsilon_1\right)_{T_{sys}} = -\left( \partial N_1/ \partial T_{sys}\right)_{\epsilon_1}
\end{equation}
that comes from the inclusion of local as well as global terms in the free energy: $N_1d\epsilon_1$ in addition to $S_{sys}dT_{sys}$, where $N_1$ and $\epsilon_1$ are the number of particles in and energy level of dot 1, whereas $T_{sys}$ and $S_{sys}$ are the temperature and entropy of the full system.
This relation is valid as long as the experimental knob that controls $\epsilon_1$ affects the energy of the thermodynamic system only through the product $N_1\epsilon_1$.   

\begin{figure}
	\includegraphics[width=\columnwidth]{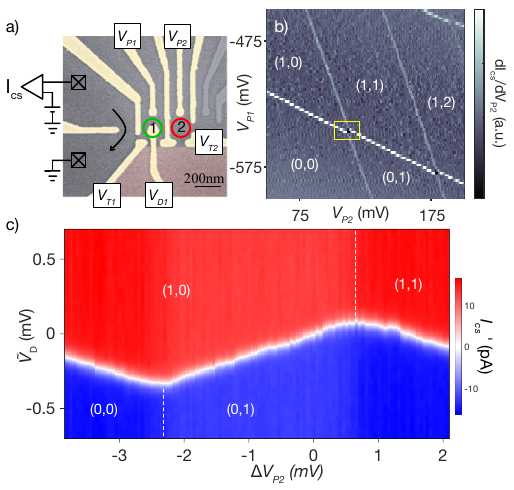}
	\caption{\label{fig:fig1} 
	a) A false colour scanning electron micrograph of the device, showing QD1 and QD2 independently tunnel coupled to a thermal reservoir, with a quantum point contact (QPC) to monitor the charge on the dots. b) The charging diagram of the double-dot system, with bright lines representing QD1 charge transitions and faint lines representing QD2 charge transitions (visible through weaker coupling of QD2 to the charge sensor). c) Adjusted charge sensor signal, $\ics'$, across the range indicated by the yellow square in (b), but collected using the virtual gate $\vd$ instead of $\vpone$.  $\ics'$ is calculated from $\ics$ by removing a slope and offset, see SI.  $(N_1,N_2)$ pairs denote regions of fixed occupation, separated by white lines. ($T=52$~mK)}
\end{figure}

Figure~\ref{fig:fig1}a shows the device: a pair of lateral few-electron QDs (QD1, QD2) defined by electrostatic gates on a GaAs/AlGaAs heterostructure, with a quantum point contact (QPC) charge sensor adjacent to QD1 whose current \ics~tracks \n.  Plungers $V_{P1}$ and $V_{P2}$ set the dot energies coarsely, $V_{D1}$ provides fine control of \eps, and $V_{T1,T2}$ set the tunnel couplings $\Gamma_{1,2}$ to a shared reservoir (pink) whose temperature is rapidly modulated by Joule heating as in Ref.~\onlinecite{child2022robust}.  Except for Figs.~\ref{fig:fig3}b and \ref{fig:fig4}, all data reported here are at moderately weak coupling, $\Gamma_{1,2}\sim 0.5\,k_B T$; lower $\Gamma$ made the device unstable.  
Figure~\ref{fig:fig1}b shows the $(N_1, N_2)$ charging diagram; bright lines are QD1 charge transitions and faint lines are QD2 transitions due to the much weaker coupling of QD2 to the charge sensor (see caption).  Diagonal motion of the charge transitions reflects cross-capacitance between QD1 and \vptwo, or QD2 and \vpone.  Likewise $V_{D1}$ has a cross-capacitance with QD2, so we construct a virtual gate \vd~as a linear combination of $V_{D1}$ and \vptwo~that tunes \eps~without affecting $\epsilon_2$ to first order.

Figure~\ref{fig:fig1}c maps the ground-state occupations across the $(0,0)\to(1,1)$ region as a function of \vd~and \vptwo.  The horizontal offset between the $(0,0)\to(0,1)$ and $(1,0)\to(1,1)$ charge lines reflects the interdot electrostatic interaction $U_{12}$: adding an electron to QD1 shifts the QD2 charge transition to more positive \vptwo.  It is this coupling that makes the charge of QD1 sensitive to entropy changes of QD2.

\begin{figure}
	 \includegraphics[width=\columnwidth]{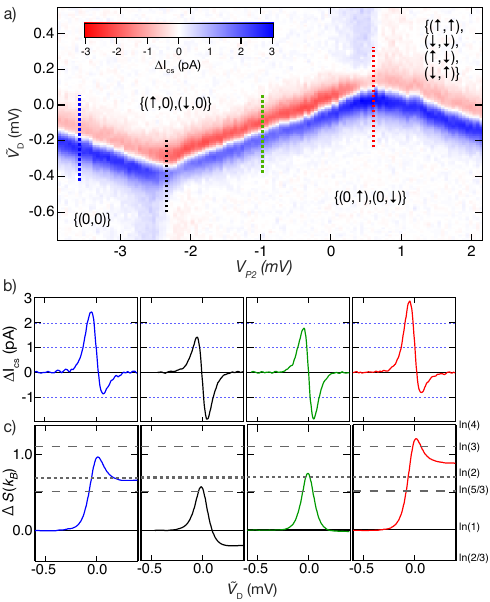}
	 \caption{\label{fig:fig2}  
	 a) Response of $\ics$, reflecting $\n$, to temperature change in the  reservoir from 52 to 73~mK, collected at the same time as Fig.~\ref{fig:fig1}c.  Regions of different QD occupations are marked, with expected spin-resolved microstates listed for each. Vertical linecuts represent trajectories in gate-space varying only $\epsilon_1$.  Data along these linecuts, b), can be integrated following Eq.~\ref{eq:eq1}, to yield $\Delta S$, shown in c).  Horizontal dotted lines represent the expected $\Delta S$ corresponding to the microstates labelled in panel a).
 	 }
\end{figure}

We now turn to the entropy measurement itself. Figure~\ref{fig:fig2}a shows the change in $I_{CS}$ induced by heating the reservoir from 52 to 73 mK, $\Delta I_{CS}\equiv I_{CS}(73~{\rm mK})-I_{CS}(52~{\rm mK})$, which is directly proportional to $\partial \n/\partial T_{sys}$ and therefore to $\partial S_{sys}/ \partial \epsilon_1$ (Eq.~\ref{eq:eq1}).  Taking advantage of the virtual gate \vd, vertical lines in Fig.~\ref{fig:fig2}a correspond to gate voltage trajectories along which only $\epsilon_1$ is varied (Figs.~\ref{fig:fig2}b).   Consider first the linecut along the blue dotted line in Fig.~\ref{fig:fig2}a, leftmost in Fig.~\ref{fig:fig2}b.  The measurement itself, $\Delta I_{CS}(\vd)$, may be converted into $\partial \n/\partial T_{sys}$ by scaling with the sensitivity of the charge sensor and the temperature change, then integrated numerically to yield $\Delta S_{sys}(\vd)$\cite{child2022robust} (Fig.~\ref{fig:fig2}c).

The resulting curve is straightforward to interpret in terms of available microstates for weakly-coupled system, enumerated for different regions of gate voltage in Fig.~\ref{fig:fig2}a \cite{Hartman.2018,child2022robust,child2022entropy}.  The system starts at the beginning of the blue dotted line with only one microstate available, $(0,0)$. By the end there are two microstates available, $(\uparrow,0)$ and $(\downarrow,0)$, corresponding to a spin-up or spin-down electron in QD1 and none in QD2, giving $\Delta S_{sys}=k_B\ln(2)$.  In the middle of the transition all three microstates are available, so the curve peaks near $\Delta S_{sys}=k_B\ln(3)$, although it does not quite reach this value due to finite dot-reservoir coupling, as discussed later in the context of Fig.~\ref{fig:fig3}.  This analysis of a double-QD structure maps directly to previous entropy measurements in single QDs because the state of QD2 does not change, and as a result, $dS_{sys}$ is the same as $dS_1$ \cite{Hartman.2018,child2022robust}.

For the rest of the trajectories through Fig.~\ref{fig:fig2}a, the available states in the double-QD device are modified by interdot interactions and the simplification $dS_{sys}=dS_1$ no longer applies. Consider, first, the green dotted linecut at $\vptwo=-1$ mV, also shown in Figs.~\ref{fig:fig2}b and \ref{fig:fig2}c. The system starts with one electron in QD2 and none in QD1, giving two possible microstates, $\{(0,\uparrow),(0,\downarrow)\}$.  Raising \vd~brings an electron into QD1, but in the process the electron is pushed out of QD2 due to mutual repulsion, so at the end of the green trajectory only the microstates $\{(\uparrow,0),(\downarrow,0)\}$ are available.  Along this trajectory, the entropy of QD1 has changed from an initial entropy $0$ to a final entropy of $k_B\ln(2)$, giving $\Delta S_1=k_B\ln(2)-0=k_B\ln(2)$,  whereas that of the whole system has changed by $\Delta S_{sys}=k_B\ln(2)-k_B\ln(2)=0$.  As expected from the thermodynamic analysis of the system, $\Delta S(\vd)$ obtained from the numerical integration along the green trajectory ends near 0, thus indicating that the measurement of $N_1$ gives access to $S_{sys}$.  Midway through the trajectory, with (0,1) and (1,0) occupations of the double dot degenerate, all four microstates $\{(0,\uparrow),(0,\downarrow),(\uparrow,0),(\downarrow,0)\}$ are available and $\Delta S_{sys}$ peaks at $k_B\ln(4)-k_B\ln(2)=k_B\ln(2)$.

In this framework, measurements of $\Delta S(\vd)$ along the black and red trajectories in Fig.~\ref{fig:fig2}a are also straightforward to interpret.  The black trajectory starts at the QD2 charge degeneracy with QD1 empty (3 microstates), passes through a 5-microstate crossing, and ends at QD1 occupied and QD2 empty (2 microstates), giving $\Delta S_\mathrm{max}=k_B\ln(5/3)$ and $\Delta S_\mathrm{1-0}=k_B\ln(2/3)$. The red trajectory starts with QD1 empty but QD2 occupied (2 microstates) and ends with QD1 occupied and QD2 at its charge transition (6 microstates), passing through the transition with all 8 microstates available so $\Delta S_\mathrm{max}=k_B\ln(8/2)$ and $\Delta S_\mathrm{1-0}=k_B\ln(6/2)$. The faint $\Delta \ics$ signal observed at QD2 transitions in Fig.~\ref{fig:fig2}a reflects a small coupling between the charge sensor and QD2 (a small sensitivity of the charge sensor to $N_2$), leading to spurious signals at the beginning and end of the black and red trajectories.  These artifacts are subtracted off, as described in the Supplementary Material, before extraction of the linecuts in Figs.~\ref{fig:fig2}b,c.

\begin{figure}
	\includegraphics[width=\columnwidth]{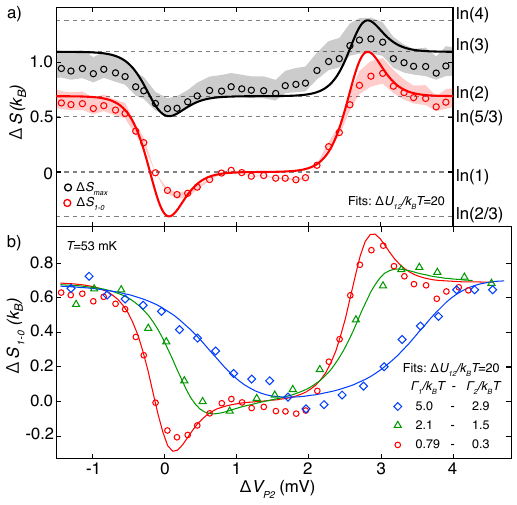}
	\caption{\label{fig:fig3}
	a) Peak (black) and final (red) measurements of entropy change across the QD1 $0\rightarrow 1$ transition, obtained by integrating vertical linecuts from Fig.~\ref{fig:fig2} after scaling to convert $\Delta I_{CS}$ to $\partial \n/\partial T$.  Solid lines are calculated entropy changes in the weak coupling limit. Shading represents the systematic uncertainty associated with imperfect knowledge of $\Delta T$ between hot and cold data. b) Net entropy change for the QD1 $0\rightarrow 1$ transition as couplings $\Gamma_{1,2}$ to the reservoir are increased.  Data match the NRG calculations well using $\Gamma_{1,2}$ extracted from independent measurements (see S.I), with only the $V_{P2}$ lever arm and a horizontal offset as fitting parameters.  The weakly-coupled dataset is the same as in panel a). Systematic uncertainties are similar to those shown in panel a). 
	}
\end{figure}

Figure~\ref{fig:fig3} shows the peak (positive maximum) and final entropy change along linecuts analogous to those shown in Fig.~\ref{fig:fig2}b, but covering the full gate space in Fig.~\ref{fig:fig2}a.  The case of moderately weak dot-reservoir coupling is shown in Fig.~\ref{fig:fig3}a.  An analytical expression can be derived in the weak coupling limit, $\Gamma_{1,2}\ll k_BT$, compared to the data at $\Gamma_{1,2}\sim 0.5k_BT$; fits to the data are shown with solid lines in Fig.~\ref{fig:fig3}a, with only the interdot coupling energy, $U_{12}$, and the gate voltage lever arm (the ratio between $\Delta V_{P2}$, in Volts, and $\Delta \epsilon_2$, in eV) as fitting parameters.  Dashed horizontal lines represent expected values of entropy change based on the various accounting of microstates listed in Fig.~\ref{fig:fig2}.

The alignment of experiment and theory across the range of $\vptwo$ shown here, covering the full $(0,0)\rightarrow(1,1)$ double-dot transition, confirms that a straightforward application of the Maxwell relations to a multi-component quantum system can indeed yield accurate measurements of entropy changes at the sub-$k_B$ level.  The experiment-theory match is significantly improved when finite dot-reservoir coupling is taken into account using numerical renormalization group (NRG) calculations  following the approach outlined in Ref.~\onlinecite{child2022entropy}.  Figure~\ref{fig:fig3}b illustrates the evolution of the net entropy change, $\Delta S_{0\rightarrow 1}$, as both dots are more and more strongly coupled to the reservoir.  The microstate-counting analysis of Fig.~\ref{fig:fig2} no longer applies, as the system must be described in terms of electron wavefunctions hybridized across dot and lead, yet the Maxwell relation determination of the entropy change for the full quantum system remains robust.  Solid lines representing NRG calculations agree closely with the data across the full range of coupling shown.  Most notable in the evolution from weak to strong coupling is the disappearance of the peak and dip from the weakly coupled data, around $\Delta\vptwo=2.8$~mV and 0.1~mV respectively.  The peak and dip features reflect the entropy associated with multiple charge states at the QD2 charge transitions; this excess entropy disappears once $\Gamma_2\gg T$ \cite{child2022entropy}, with the various charge states of QD2 strongly hybridized by $\Gamma_2$.

Figure~\ref{fig:fig4} explores the entropy change, $\Delta S(\vd)$, across the $(0,1)$--$(1,0)$ transition for the most strongly-coupled device setting in Fig.~\ref{fig:fig3}b.  The data are collected midway between the triple points, as far as possible from the gate voltage settings where single-dot transitions are allowed.  In our capacitively-coupled geometry, the $(0,1)$--$(1,0)$ degeneracy cannot be lifted by direct interdot tunneling, but is instead lifted by virtual cotunneling via the reservoir, giving a hybridization scale of order $4\Gamma_1\Gamma_2/U_{12}$.  For the data in Fig.~\ref{fig:fig4}, this scale is of order $100$~mK.   In the high-temperature limit, the peak in $\Delta S$ exceeds $\kb\ln 2$ then settles to a final value above zero, likely reflecting excess entropy from the single dot QD2 transition. As $T$ decreases below the hybridization scale, the peak is progressively suppressed, analogous to the collapse in single dot entropy at the charge transition due to dot-lead coupling in Ref.~\onlinecite{child2022entropy}.   Because all relevant energy scales---$\Gamma_{1,2}$, $U_{12}$, and $k_B T$---are of the same order in the device of Fig.~\ref{fig:fig4}, the entropy contributions from the two dots overlap and cannot be cleanly separated, limiting the comparison with NRG to the qualitative level of the suppression trend.

\begin{figure}
	\includegraphics[width=\columnwidth]{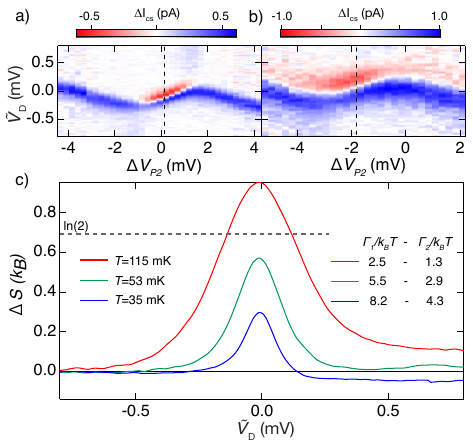}
	\caption{\label{fig:fig4}
	a,b) Maps of $dN/dT$ analogous to Fig.~\ref{fig:fig2}a, for the strongest-coupled setting in Fig.~\ref{fig:fig3}, at $T=35$, $115$~mK for panels a, b respectively. c) Measured $\Delta S(\vd)$ across the (0,1)-(1,0) transition along the linecuts marked in a,b) chosen to be midway between the triple points, as well as an equivalent dataset at 52~mK, showing the collapse in entropy associated with the onset of interdot correlations via the reservoir. 
	}
\end{figure}

In summary, our results demonstrate that Maxwell-relation measurements provide a robust probe of entropy changes in coupled quantum systems, even when the particles whose charge is measured are not themselves the carriers of the entropy being sensed.  In this way, the quantum dot functions as an effective entropy sensor, transducing entropy variations in a nearby circuit element into measurable charge signals.  This capability opens the door to the use of entropic characterization as a means to identify exotic quasiparticles, beyond electrons, that may emerge in mesoscopic quantum circuits.

ACKNOWLEDGEMENTS: The authors acknowledge helpful discussions with A. Mitchell.  This project has received funding from European Research Council (ERC) under the European Union’s Horizon 2020 research and innovation program under grant agreement No 951541. Y. Meir acknowledges support by the Israel Science Foundation (grant 3523/2020).  Experiments at UBC were undertaken with support from the Stewart Blusson Quantum Matter Institute, the Natural Sciences and Engineering Research Council of Canada, the Canada Foundation for Innovation, the Canadian Institute for Advanced Research, and the Canada First Research Excellence Fund, Quantum Materials and Future Technologies Program.  Work in the Manfra group at Purdue University was supported by the US DOE Office of Basic Energy Sciences under Award DE-SC0020138.

\bibliography{main}

@article{galpin2005quantum,
  title={Quantum phase transition in capacitively coupled double quantum dots},
  author={Galpin, Martin R and Logan, David E and Krishnamurthy, HR},
  journal={Physical review letters},
  volume={94},
  number={18},
  pages={186406},
  year={2005},
  publisher={APS}
}

@article{chan2002strongly,
  title={Strongly capacitively coupled quantum dots},
  author={Chan, Ian H and Westervelt, RM and Maranowski, KD and Gossard, AC},
  journal={Applied Physics Letters},
  volume={80},
  number={10},
  pages={1818--1820},
  year={2002},
  publisher={American Institute of Physics}
}

@article{hubel2007two,
  title={Two laterally arranged quantum dot systems with strong capacitive interdot coupling},
  author={H{\"u}bel, A and Weis, J and Dietsche, W and Klitzing, K v},
  journal={Applied Physics Letters},
  volume={91},
  number={10},
  year={2007},
  publisher={AIP Publishing}
}

@article{krychowski2016spin,
  title={Spin-orbital and spin Kondo effects in parallel coupled quantum dots},
  author={Krychowski, D and Lipi{\'n}ski, S},
  journal={Physical Review B},
  volume={93},
  number={7},
  pages={075416},
  year={2016},
  publisher={APS}
}

@article{mitchell2006gate,
  title={Gate voltage effects in capacitively coupled quantum dots},
  author={Mitchell, Andrew K and Galpin, Martin R and Logan, David E},
  journal={Europhysics Letters},
  volume={76},
  number={1},
  pages={95},
  year={2006},
  publisher={IOP Publishing}
}

@article{PhysRevB.84.161305,
  title = {Spin-orbital Kondo effect in a parallel double quantum dot},
  author = {Okazaki, Yuma and Sasaki, Satoshi and Muraki, Koji},
  journal = {Phys. Rev. B},
  volume = {84},
  issue = {16},
  pages = {161305},
  numpages = {4},
  year = {2011},
  month = {Oct},
  publisher = {American Physical Society},
  doi = {10.1103/PhysRevB.84.161305},
  url = {https://link.aps.org/doi/10.1103/PhysRevB.84.161305}
}

@article{PhysRevLett.101.186804,
  title = {Correlated Electron Tunneling through Two Separate Quantum Dot Systems with Strong Capacitive Interdot Coupling},
  author = {H\"ubel, A. and Held, K. and Weis, J. and v. Klitzing, K.},
  journal = {Phys. Rev. Lett.},
  volume = {101},
  issue = {18},
  pages = {186804},
  numpages = {4},
  year = {2008},
  month = {Oct},
  publisher = {American Physical Society},
  doi = {10.1103/PhysRevLett.101.186804},
  url = {https://link.aps.org/doi/10.1103/PhysRevLett.101.186804}
}

@article{PhysRevLett.110.046604,
  title = {Pseudospin-Resolved Transport Spectroscopy of the Kondo Effect in a Double Quantum Dot},
  author = {Amasha, S. and Keller, A. J. and Rau, I. G. and Carmi, A. and Katine, J. A. and Shtrikman, Hadas and Oreg, Y. and Goldhaber-Gordon, D.},
  journal = {Phys. Rev. Lett.},
  volume = {110},
  issue = {4},
  pages = {046604},
  numpages = {5},
  year = {2013},
  month = {Jan},
  publisher = {American Physical Society},
  doi = {10.1103/PhysRevLett.110.046604},
  url = {https://link.aps.org/doi/10.1103/PhysRevLett.110.046604}
}

@article{keller2014emergent,
  title={Emergent SU (4) Kondo physics in a spin--charge-entangled double quantum dot},
  author={Keller, AJ and Amasha, S and Weymann, Ireneusz and Moca, CP and Rau, IG and Katine, JA and Shtrikman, Hadas and Zar{\'a}nd, G and Goldhaber-Gordon, D},
  journal={Nature Physics},
  volume={10},
  number={2},
  pages={145--150},
  year={2014},
  publisher={Nature Publishing Group UK London}
}

@article{hou2020many,
  title={Many-body tunneling and nonequilibrium dynamics in double quantum dots with capacitive coupling},
  author={Hou, Wenjie and Wang, Yuandong and Zhao, Weisheng and Zhu, Zhengang and Wei, Jianhua and Luo, Honggang and Yan, Yijing},
  journal={Journal of Physics: Condensed Matter},
  volume={33},
  number={7},
  pages={075301},
  year={2020},
  publisher={IOP Publishing}
}

@article{lombardo2020kondo,
  title={Kondo-assisted switching between three conduction states in capacitively coupled quantum dots},
  author={Lombardo, Pierre and Hayn, Roland and Zhuravel, Denis and Sch{\"a}fer, Steffen},
  journal={Physical Review Research},
  volume={2},
  number={3},
  pages={033387},
  year={2020},
  publisher={APS}
}

@article{ferreira2011capacitively,
  title={Capacitively coupled double quantum dot system in the Kondo regime},
  author={Ferreira, Irisnei L and Orellana, PA and Martins, GB and Souza, FM and Vernek, E},
  journal={Physical Review B—Condensed Matter and Materials Physics},
  volume={84},
  number={20},
  pages={205320},
  year={2011},
  publisher={APS}
}

@article{borda20034,
  title={SU (4) Fermi liquid state and spin filtering in a double quantum dot system},
  author={Borda, L{\'a}szl{\'o} and Zar{\'a}nd, Gergely and Hofstetter, Walter and Halperin, BI and von Delft, Jan},
  journal={Physical review letters},
  volume={90},
  number={2},
  pages={026602},
  year={2003},
  publisher={APS}
}

@article{pyurbeeva2021controlling,
  title={Controlling the entropy of a single-molecule junction},
  author={Pyurbeeva, Eugenia and Hsu, Chunwei and Vogel, David and Wegeberg, Christina and Mayor, Marcel and Van Der Zant, Herre and Mol, Jan A and Gehring, Pascal},
  journal={Nano Letters},
  volume={21},
  number={22},
  pages={9715--9719},
  year={2021},
  publisher={ACS Publications}
}

@article{Kealhofer2025Entropy,
  title = {Entropy of a Double Quantum Dot},
  author = {Kealhofer, David and Adam, Christoph and Ruckriegel, Max J. and Tomi\ifmmode \acute{c}\else \'{c}\fi{}, Petar and Kratochwil, Benedikt and Reichl, Christian and Meir, Yigal and Wegscheider, Werner and Ihn, Thomas and Ensslin, Klaus},
  journal = {Phys. Rev. Lett.},
  volume = {135},
  issue = {20},
  pages = {206303},
  numpages = {5},
  year = {2025},
  month = {Nov},
  publisher = {American Physical Society},
  doi = {10.1103/gl59-td1w},
  url = {https://link.aps.org/doi/10.1103/gl59-td1w}
}

@article{Adam2025Entropy,
  title = {Entropy Spectroscopy of a Bilayer Graphene Quantum Dot},
  author = {Adam, C. and Duprez, H. and Lehmann, N. and Yglesias, A. and Denisov, A. O. and Cances, S. and Ruckriegel, M. J. and Masseroni, M. and Tong, C. and Huang, W. and Kealhofer, D. and Garreis, R. and Watanabe, K. and Taniguchi, T. and Ensslin, K. and Ihn, T.},
  journal = {Phys. Rev. Lett.},
  volume = {135},
  issue = {12},
  pages = {126202},
  numpages = {6},
  year = {2025},
  month = {Sep},
  publisher = {American Physical Society},
  doi = {10.1103/vbbj-138r},
  url = {https://link.aps.org/doi/10.1103/vbbj-138r}
}

@article{child2022robust,
  title={A robust protocol for entropy measurement in mesoscopic circuits},
  author={Child, Timothy and Sheekey, Owen and L{\"u}scher, Silvia and Fallahi, Saeed and Gardner, Geoffrey C and Manfra, Michael and Folk, Joshua},
  journal={Entropy},
  volume={24},
  number={3},
  pages={417},
  year={2022},
  publisher={MDPI}
}

@article{child2022entropy,
  title={Entropy measurement of a strongly coupled quantum dot},
  author={Child, Timothy and Sheekey, Owen and L{\"u}scher, Silvia and Fallahi, Saeed and Gardner, Geoffrey C and Manfra, Michael and Mitchell, Andrew and Sela, Eran and Kleeorin, Yaakov and Meir, Yigal and others},
  journal={Physical Review Letters},
  volume={129},
  number={22},
  pages={227702},
  year={2022},
  publisher={APS}
}

@article{saito2021isospin,
  title={Isospin Pomeranchuk effect in twisted bilayer graphene},
  author={Saito, Yu and Yang, Fangyuan and Ge, Jingyuan and Liu, Xiaoxue and Taniguchi, Takashi and Watanabe, Kenji and Li, JIA and Berg, Erez and Young, Andrea F},
  journal={Nature},
  volume={592},
  number={7853},
  pages={220--224},
  year={2021},
  publisher={Nature Publishing Group UK London}
}

@article{sankar2023measuring,
  title={Measuring topological entanglement entropy using Maxwell relations},
  author={Sankar, Sarath and Sela, Eran and Han, Cheolhee},
  journal={Physical Review Letters},
  volume={131},
  number={1},
  pages={016601},
  year={2023},
  publisher={APS}
}

@article{Sela.2019, 
year = {2019}, 
title = {{Detecting the universal fractional entropy of Majorana zero modes}}, 
author = {Sela, Eran and Oreg, Yuval and Plugge, Stephan and Hartman, Nikolaus and Lüscher, Silvia and Folk, Joshua}, 
journal = {arXiv}, 
doi = {10.1103/physrevlett.123.147702}, 
pmid = {31702171}, 
eprint = {1905.12237}
}

@article{Hartman.2018, 
year = {2018}, 
title = {{Direct entropy measurement in a mesoscopic quantum system}}, 
author = {Hartman, Nikolaus and Olsen, Christian and Lüscher, Silvia and Samani, Mohammad and Fallahi, Saeed and Gardner, Geoffrey C. and Manfra, Michael and Folk, Joshua}, 
journal = {Nature Physics}, 
issn = {1745-2473}, 
doi = {10.1038/s41567-018-0250-5}, 
eprint = {1905.12388}, 
pages = {1083--1086}, 
number = {11}, 
volume = {14}, 
}

@article{Cooper.2008, 
year = {2008}, 
title = {{Observable Bulk Signatures of Non-Abelian Quantum Hall States}}, 
author = {Cooper, N R and Stern, Ady}, 
journal = {Physical Review Letters}, 
doi = {10.1103/physrevlett.102.176807}, 
pmid = {19518812}, 
eprint = {0812.3387}, 
}

@article{Rozen.2021, 
year = {2021}, 
title = {{Entropic evidence for a Pomeranchuk effect in magic-angle graphene}}, 
author = {Rozen, Asaf and Park, Jeong Min and Zondiner, Uri and Cao, Yuan and Rodan-Legrain, Daniel and Taniguchi, Takashi and Watanabe, Kenji and Oreg, Yuval and Stern, Ady and Berg, Erez and Jarillo-Herrero, Pablo and Ilani, Shahal}, 
journal = {Nature}, 
issn = {0028-0836}, 
doi = {10.1038/s41586-021-03319-3}, 
pmid = {33828314}, 
eprint = {2009.01836}, 
pages = {214--219}, 
number = {7853}, 
volume = {592}, 
}

@article{Pascal.2021,
  title={Complete mapping of the thermoelectric properties of a single molecule},
  author={Gehring, Pascal and Sowa, Jakub K and Hsu, Chunwei and de Bruijckere, Joeri and van der Star, Martijn and Le Roy, Jennifer J and Bogani, Lapo and Gauger, Erik M and van der Zant, Herre SJ},
  journal={Nature nanotechnology},
  volume={16},
  number={4},
  pages={426--430},
  year={2021},
  publisher={Nature Publishing Group UK London}
}

@article{BenShach.2013, 
year = {2013}, 
title = {{Detecting Non-Abelian Anyons by Charging Spectroscopy}}, 
author = {Ben-Shach, G. and Laumann, C. R. and Neder, I. and Yacoby, A. and Halperin, B. I.}, 
journal = {Physical Review Letters}, 
issn = {0031-9007}, 
doi = {10.1103/physrevlett.110.106805}, 
pmid = {23521280}, 
eprint = {1212.1163}, 
pages = {106805}, 
number = {10}, 
volume = {110}
}
\newpage
\clearpage

\onecolumngrid

\renewcommand{\figurename}{Supplementary Data Fig.}
\renewcommand{\thesubsection}{S\arabic{subsection}}
\setcounter{secnumdepth}{2}
\setcounter{figure}{0} 
\setcounter{equation}{0}

\onecolumngrid

\section*{Supplementary Information}

\subsection{Technical notes on data acquisition}

Gate voltages and heater bias currents were set by DACs operating at a sampling rate of 2531Hz. The charge sensor current was read out via a current-to-voltage converter using an ADC synchronized to the DAC steps.  The synchronized DAC/ADC units were optimized by our group and the UBC Physics and Astronomy technical staff, starting from the OpenDAC platform (https://opendacs.com/) developed by Hugh Churchill and Andrea Young.  Following the open-source approach from Young and Churchill, we are happy to share technical details with interested parties.

The data collection process is described in detail in Ref.~\cite{child2022robust}.  In summary, the heater current was switched rapidly between positive/zero/negative/zero values to heat/cool/heat/cool the thermal reservoir [pink in Fig.~1(a)] with a periodic three-level periodic waveform  applied to the bias resistor at the heater source contact, and a synchronized periodic voltage applied to the heater drain in order to hold the heated reservoir at zero potential.  This heater cycle was carried out at each value of gate voltage, then gate(s) were stepped and the heating cycle was repeated again.  Each hot or cold step lasted 20~ms (51 DAC/ADC samples at 2531Hz).  When only the heater current changed between steps, the first 4 points were removed to allow the circuit to reach equilibrium.  After completing a hot/cold/hot/cold cycle, the gate would be stepped, and the first 15 points were removed to account for spurious signals due to the gate voltage step.

\subsection{Processing notes for Fig. 1}

\noindent \textbf{Figure~1(b)} represents a scan taken by sweeping over $V_{P2}$, stepping $V_{P1}$.  The collected charge sensor current $I_{CS}$ is then numerically differentiated with respect to $V_{P2}$.\\

\begin{figure*}[!ht]
    \centering\includegraphics{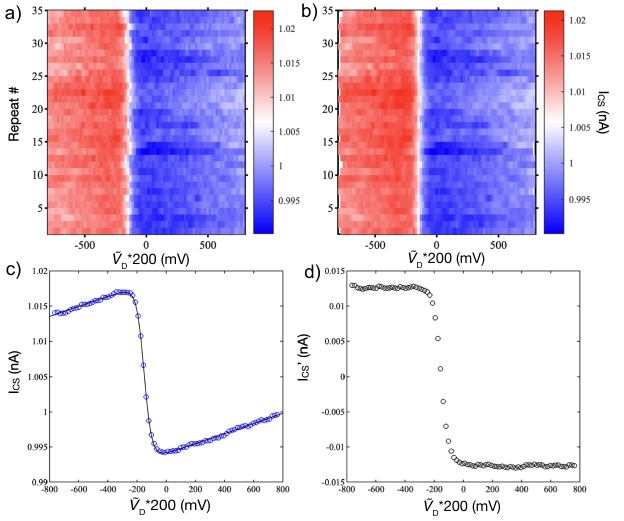}
    \caption{This figure illustrates how the data in Fig.~\ref{fig:fig1}(c) from the main text were processed.  At each step in $V_{P2}$, multiple scans of $\vd$ were collected.  Panel a) shows an example with 35 repeated scans.  The factor of 200 in the x-axis label reflects a 200 times voltage divider in the circuit.  Then, the charge transition lineshape was fit to each line allowing for an offset in the x-axis, and the individual traces were shifted horizontally to center each at the average offset value.  The shifted data are shown in b).  These 35 shifted traces were then averaged together to yield the data in c), which shows the charge transition lineshape fit to the averaged data.  The slope away from the charge transition reflects a linear background included in the fit to allow for cross capacitance between $\vd$ and the charge sensor.  This linear background was subtracted off to yield the adjusted charge sensor current $I_{CS}'$ in d).} 
    \label{sfig:1}
\end{figure*}

\noindent \textbf{Figure~1(c)} represents sweeps over the virtual gate $\vd$, stepping $V_{P2}$. $\vd$ is itself composed of a linear combination of $V_{D1}$ and a correction to $V_{P2}$ that counteracts (to first order) the cross-capacitive influence of $V_{D1}$ on dot 2. In practice this is done by using a voltage divider circuit to add a correction voltage, generated by separate DAC channel and calculated to be proportional to $V_{D1}$, to the voltage $V_{P2}$ before the signal is sent to the gate.  In order to reduce measurement noise,  multiple $\vd$ sweeps were performed at each $V_{P2}$, then shifted and averaged together and adjusted to remove a linear background as described in Supp.~Fig.~\ref{sfig:1} (see caption for details).   These adjusted $I_{CS}'$ data, at each value of $V_{P2}$, were then assembled into the 2D data in Fig.~\ref{fig:fig1}c.

In practice, the data in Fig.~1(c) and Fig.~2(a) were collected effectively at the same time.  
Fig.~1(c) represents a `cold' transition, and Fig.~2(a) represents the difference between hot and cold transitions.  Starting from the hot/cold/hot/cold data collected at each value of gate voltage (see section S1), hot and cold data were separated then averaged to yield a single `hot' and `cold' signal at each gate voltage, corresponding to base temperature ($T$) and heated ($T+dT$) occupation data.  For Fig.~1(c), we show only the occupation data collected in the cold state, that is, at the sample temperature: 52 mK in this case.  An advantage of collecting the hot and cold data together is that the shifting procedure described in Supp.~Fig.~\ref{sfig:1} can be done without corrupting the subtraction of hot and cold signals. 

\subsection{Processing notes for Fig. 2}

\noindent \textbf{Figure~\ref{fig:fig2}a} represents the difference in occupation due to a 21$\pm$2~mK increase in temperature, from 52 mK as shown in Fig.~1(c) to 73~mK. As described in the section above, the 52 and 73~mK data are collected in an interlaced fashion, then subtracted after processing.  Because the hot and cold data are collected nearly simultaneously, any shifts in the dot energies due to charge noise in the device are removed and do not create a spurious signal.\\

\noindent In order to extract the data in \textbf{Fig.~\ref{fig:fig2}b}, from which $\Delta S$ may be calculated, further averaging is done.  Each of the traces in Fig.~\ref{fig:fig2}b represents an average over $\pm$0.05~mV in $\vd$, including data from the 2D scan in Fig.~\ref{fig:fig2}a in addition to three analogous 2D datasets collected days or weeks later at the same gate voltage settings.  In order to account for small gate voltage shifts over that time, 2D cross-correlations between the 2D datasets were performed and the data shifted in $\vd$ and $V_{P2}$ by the location of the peak in the cross correlation.

The charge sensor in this experiment was much more strongly coupled to QD1 than to QD2.  As discussed on the main text, however, there was also a weak coupling to QD2 (that is, the charge sensor was weakly sensitive to $N_2$), which gave rise to a spurious signal during entropy measurements we needed to exclude from the data processing.  The origin of this spurious signal may be seen mathematically by considering additional terms in the thermodynamic potential used for the Maxwell relation.  Equation~\ref{eq:eq1} in the main text was derived from the $SdT$ and $N_1d\epsilon_1$ terms in the thermodynamic potential.\cite{child2022entropy} But there is also a term $N_2d\epsilon_2$ that reflects the energy of an electron in $QD_2$, which leads to another Maxwell relation, $\left(\partial S_{sys}/ \partial \epsilon_2\right)_{T_{sys}} = -\left( \partial N_2/ \partial T_{sys}\right)_{\epsilon_2}$.  In other words, when $S_{sys}$ changes with $\epsilon_2$, $N_2$ will change with $T$, and that can yield a small $T$-dependent signal in the charge detector.  Because this contribution reflects the entropy dependence on $\epsilon_2$, it must not be included in the integration along $\epsilon_1$ in order to extract a net entropy change.  In the experiment, this spurious signal can be seen as the faint blue and red vertical stripes in Fig.~2a at  $V_{P2}=-2.5$~mV and 1.5~mV, where QD2 changes occupation.  The removal process is outlined in Fig.~S2, and described in the caption.

\subsection{Calculations of $\Delta S$ in the weakly coupled limit (Fig.~3a)}

We start from the grand-canonical partition function of a capacitively coupled double quantum dot restricted to the four charge states
$(0,0)$, $(1,0)$, $(0,1)$, and $(1,1)$, where $\varepsilon_{1/2}$ are the energies to add one electron to dots 1 and 2 respectively, and $U_{12}$ is the interdot interaction as described in the main text. The prefactors for each term reflect spin degeneracy. We reference energies with respect to the $(0,0)$ state:
\[
Z =
1
+ 2 e^{-\frac{\varepsilon_1-\mu}{k_B T}}
+ 2 e^{-\frac{\varepsilon_2-\mu}{k_B T}}
+ 4 e^{-\frac{\varepsilon_1+\varepsilon_2+U_{12}-2\mu}{k_B T}}.
\]

Using $\langle n \rangle=\frac{1}{Z}\sum_i n_i e^{-E_i/(k_B T)}$, we then have

\[
\frac{\partial n_1}{\partial T}
=
\frac{
2 e^{\frac{\mu-\varepsilon_1}{k_B T}}(\varepsilon_1-\mu)
+ 4 e^{\frac{2\mu-\varepsilon_1-\varepsilon_2}{k_B T}}(\varepsilon_1-\varepsilon_2)
+ 4 e^{\frac{2\mu-\varepsilon_1-\varepsilon_2-U_{12}}{k_B T}}(\varepsilon_1+\varepsilon_2+U_{12}-2\mu)
+ 8 e^{\frac{3\mu-\varepsilon_1-2\varepsilon_2-U_{12}}{k_B T}}(\varepsilon_1+U_{12}-\mu)
}{
Z^2 (k_B T)^2
}
\]

From this, $S(\varepsilon_1)$ is calculated by integration of the Maxwell relation.

\subsection{Estimating $\Gamma_{1,2}$}

In order to estimate values of $\Gamma_{1,2}$ in this experiment, charge transition data for the independent dots (0,0)-(1,0) and (0,0)-(0,1) were collected for varying $V_{T1,T2}$, going from $\Gamma<T$ to $\Gamma>T$ where $T=52$~mK.  Then, the data were fit simultaneously to NRG simulations covering a range of $\Gamma/T$, allowing independent $\Gamma$ for each charge transition.  Because $T$ was known, $\Gamma$ could then be determined.  We note that this approach assumes a fixed lever-arm, independent of $V_T$. 

\subsection{Estimating hybridization energies}

Our estimate of the scale on which virtual cotunneling mixes the $(0,1)$ and $(1,0)$ states is obtained from a spinless four-state model of the capacitively coupled double dot in the basis $\{|0,0\rangle,|0,1\rangle,|1,0\rangle,|1,1\rangle\}$, with dot-lead matrix elements $\Gamma_{1,2}$, interdot electrostatic interaction energy $U_{12}$, and no direct interdot tunneling.  At the midpoint of the $(0,1)$--$(1,0)$ degeneracy line, the virtual $(0,0)$ and $(1,1)$ intermediates each lie at energy $U_{12}/2$; second-order perturbation theory then gives an effective hopping $4\Gamma_1\Gamma_2/U_{12}$ between the two charge configurations, hence a bonding-antibonding splitting of the same parametric form.  The corresponding prefactor in the spinful experiment requires an NRG treatment and is not attempted here, since Fig.~\ref{fig:fig4} is interpreted only at a qualitative level.

\newpage

\subsection{Supplement Figures}

\begin{figure*}[!ht]
    \centering\includegraphics{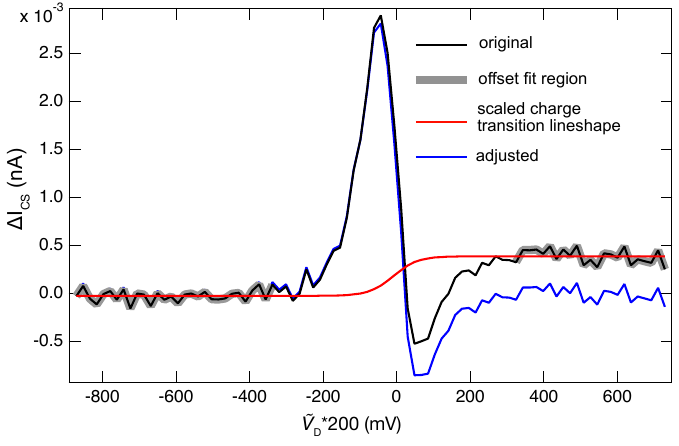}
    \caption{This figure describes the subtraction of spurious signals in $\Delta I_{CS}(\vd)$ away from the QD1 charge transition--for example those due to small but non-zero coupling of the charge sensor to QD2--before further analysis.   The data shown here (grey) represent a single vertical slice from 2D data like those shown in Fig.~\ref{fig:fig2}a.  Regions well to the left and to the right of the charge transition itself (thicker black, corresponding to $N_1=0$ and $N_1=1$ for left and right respectively) are averaged to find left and right offsets (nearly zero on the left, in this case, and around 0.25~pA on the right).  A function (red) proportional to the charge transition itself, extracted from $I_{CS}(\vd)$, but scaled to transition from left to right offsets, is subtracted off the raw (grey) data yielding the blue trace, which is then used for further analysis.  The motivation to scale the charge transition for offset subtraction is that the difference in offset between left and right regions apparently depends on the QD1 charge, so to a first approximation the offset should shift between the two values in proportion to the charge on QD1.} 
    \label{sfig:2}
\end{figure*}

\end{document}